\documentclass{article}

\usepackage[applemac]{inputenc}

\usepackage{graphicx} 
\usepackage{amsmath}
\usepackage{amsfonts}
\usepackage{amssymb}
\usepackage{dsfont}
\usepackage[english]{babel}
\usepackage[authoryear,round]{natbib}
\usepackage{rotating}
\usepackage{booktabs}
\usepackage{stackrel}

\title{A new time-varying model for forecasting long-memory series}
\author{Luisa Bisaglia and Matteo Grigoletto\\
\small Department of Statistical Sciences\\
\small University of Padova\\
\footnotesize\texttt{\{luisa.bisaglia, matteo.grigoletto\}@unipd.it}}
\date{}

\begin{document}
\maketitle

\begin{abstract}
\noindent In this work we propose a new class of long-memory models with
time-varying fractional parameter. In particular, the dynamics of the
long-memory coefficient, $d$, is specified through a stochastic
recurrence equation driven by the score of the predictive likelihood,
as suggested by \cite{creal:2013} and \cite{harvey:2013}. We
demonstrate the validity of the proposed model by a Monte Carlo
experiment and an application to two real time series.
\end{abstract}
{\bf Keywords:} long-memory, GAS model, time-varying parameter.

\section{Introduction}

Long-memory processes have proved to be useful tools in the
analysis of many empirical time series. 
These series present the property that the autocorrelation function at
large lags decreases to zero like a power function rather than
exponentially, so that the correlations are not summable. 
In the frequency domain, this means that the spectral density behaves
like a power function and it diverges as the frequencies tend to
zero.\\
One of the most popular processes that takes into account this
particular behavior of the autocorrelation function is the
AutoRegressive Fractionally Integrated
Moving Average process (ARFIMA$(p,d,q)$), independently
introduced by \cite{granger:1980} and \cite{hosking:1981}. This process
generalizes the ARIMA$(p,d,q)$ process by relaxing the assumption
that $d$ is an integer.\\
The ARFIMA$(p,d,q)$ process,
$Y_t,$
is defined by the difference equation
$$\Phi(B)(1-B)^d (Y_t - \mu) = \Theta(B)\,\varepsilon_t,$$
where $\varepsilon_t \sim WN(0, \sigma^2),$ and $\Phi(\cdot)$ and
$\Theta(\cdot)$ are polynomials in the backward shift operator $B$ of
degrees $p$ and $q$, respectively. Furthermore, 
$(1-B)^d=\sum_{j=0}^{\infty} \pi_j B^j$,
with $\pi_j = \Gamma(j-d)/(\Gamma(j+1)\Gamma(-d))$,
where $\Gamma(\cdot)$ denotes the gamma function.
When the roots of $\Phi(B)=0$ and $\Theta(B)=0$
lie outside the unit circle and $\mid d \mid <0.5$, the process is
stationary, causal and invertible. We will assume these conditions
to be satisfied.\\
When $d\in(0,0.5)$ the autocorrelation function of the process decays
to zero hyperbolically at a rate $O(k^{2d-1})$, where $k$ denotes the
lag. In this case we say that the process has a long-memory
behavior. When $d\in(-0.5,0)$ the process is said to have intermediate
memory.

If $p=q=0$, the process $\{Y_t,\;t = 0,\pm 1,\ldots\}$
is called Fractionally Integrated Noise, FI$(d)$. In the following we
will concentrate on FI$(d)$ processes with $d\in(-0.5,0.5)$.
%We will also assume, for convenience and without loss of generality,
%that $\mu=0$ and $\sigma^2=1$.

Several papers have addressed the detection of breaks in the order of
fractional integration. Some of these works allowed for just one
unknown breakpoint (see, for instance, \citealp{beran:1996};
\citealp{yamaguchi:2011}).
Others treated the number of breaks as well as their timing as unknown
(\citealp{ray:2002}; \citealp{hasller:2014}). \cite{boutahar:2008}
and, more recently, \cite{boubaker:2018}
generalize the standard long-memory modeling by assuming that the
long-memory parameter $d$ is stochastic and time-varying. The authors
introduce a STAR process, characterized by a logistic
function, on this parameter and propose an estimation method for the
model. \citet{Caporin:2013} propose a variation of the ARFIMA model,
allowing for monthly changes in the memory coefficient through a step
function. Finally, \citet{jensen2000}, \citet{roueff:2011} and
\citet{lu:2011}  take into account the time-varying feature of the
long-memory parameter $d$ using the wavelets approach.

Our approach is completely different because we allow the long-memory
parameter $d$ to vary at each time $t$. Moreover, our approach is
based on the theory of Generalized Autoregressive Score (GAS)
models. In particular, the peculiarity of our approach is that the
dynamics of the long-memory parameter is specified through a stochastic
recurrence equation driven by the score of the predictive
likelihood. In this way we are able to take into account also smooth
changes of the long-memory parameter. 
%Thus comparison with the previous mentioned methods are not possible,
%because our method allows to estimate the whole trajectory of $d_t$
%not only two or three regimes.

The paper is organized as follows. Section~\ref{sec:gasmodel} recalls
GAS models. In Section~\ref{sec:tvfi} our time-varying long-memory
model is proposed and the maximum likelihood estimation procedure is
introduced. Section~\ref{sec:mc} reports the results of some Monte
Carlo experiments to evaluate the performance of the proposed
methodology. Section~\ref{sec:empapp} contains two empirical
application and Section~\ref{sec:concl} concludes.

\section{GAS model}
\label{sec:gasmodel}
To allow for time-varying parameters, \cite{creal:2013} and
\cite{harvey:2013} proposed an updating equation where the
innovation is given by the score of the conditional distribution of
the observations (GAS models). The basic framework is
the following. Consider a time series $\left\{y_1, \cdots, y_n\right\}$ with
time-$t$ observation density $p(y_t\mid\psi_t, \mathcal{F}_{t})$,
where $\psi_t=(f_t, \theta)$ is the parameter vector, with $f_t$
representing the time-varying parameter(s) and $\theta$ the remaining
fixed coefficients. $\mathcal{F}_{t}=\{y_1, \ldots,y_{t-1};f_1,
\ldots,f_{t-1} \}$ is the available information set at time $t$.

In time series the likelihood function can be written via prediction
errors as:
$$\mathcal{L}(y,\psi)=p(y_1;\psi_1) \prod_{t=2}^{n}p(y_t \mid
\psi_t, \mathcal{F}_t)\ .$$
Thus, the $t$-th contribution to the log-likelihood is:
$$l_t=\log p(y_t \mid y_1,\cdots, y_{t-1}; f_1, \cdots, f_t; \theta)=
\log p(y_t \mid y_1,\cdots, y_{t-1}; f_t; \theta)\ ,$$
where we assume that $f_1, \cdots, f_t$ are known (because they are
realized).

The parameter value for the next period, $f_{t+1}$, is determined by
an autoregressive updating function that has an innovation equal to
the score of $l_t$ with respect to $f_t.$ In particular, when a new
observation $y_t$ is realized, we update the time-varying parameter
$f_t$ to the next period $t+1,$ assuming that:
$$f_{t+1}=\omega + \beta f_t + \alpha s_t\ ,$$
where the innovation $s_t$ is given by
$$s_t=S_t \cdot \nabla_t\ ,$$
with
\begin{equation}
\label{eq:Score_pred}
\nabla_t=\frac{\partial \log p(y_t \mid y_1, \cdots, y_{t-1}; f_t, \theta)}{\partial f_t}
\end{equation}
and $S_t=S(t,f_t,\mathcal{F}_t;\theta),$ a scaling matrix that depends
on the variance of the score. In our work, following the suggestion of
\cite{creal:2013}, we define $S_t$ as:
\begin{equation}
\label{eq:Fish_inf}
S_t=\mathcal{I}_{t-1}^{-1}=
-E_{t-1}\left[ \frac{\partial^2 \log p(y_t \mid y_1, \cdots, y_{t-1};
	f_t, \theta)}{\partial f_t \partial f_t^{'}}\right]^{-1}\ .
\end{equation}
By determining $f_{t+1}$ in this way, we obtain a recursive algorithm
for the estimation of time-varying parameters.

\section{A time-varying long-memory model}
\label{sec:tvfi}
In this section, we extend the class of FI$(d)$ models, by allowing the
long-memory parameter $d$ to change over time. The dynamics of the
time-varying coefficient $d_t$ is specified in the GAS framework
outlined above.\footnote{Note that the model we propose is different
  from the fractionally integrated GAS model, proposed in
  \cite{creal:2013}, which assumes that the updating mechanism for
  $f_t$ is given  by a long-memory model.}

The TV-FI model is described by the following equations:
\[
(1-B)^{d_t} \; y_t = \varepsilon_t\ ,
\]

\begin{equation}
\label{eq:tvid_rec}
d_{t+1} = \omega + \beta\, d_{t} + \alpha\, s_t\ ,
\end{equation}
where $\varepsilon_t \sim iid \mathcal{N}(0, \sigma^2),$ and
$s_t=S_t \nabla_t$ with $S_t$ and $\nabla_t$ defined below.

The idea behind equation~(\ref{eq:tvid_rec}) is that in some periods
the data could be more informative than in others. Suppose, for
instance, that $d_t$ has two regimes, $d_1$ for the first $\tau\, n$
and $d_2$ for the last $(1-\tau)\, n$ observations, where $n$ is the
length of the series and $\tau \in (0,1)$. Before the change, the
magnitude of the innovations $s_t$ should be small. However, after the
change new observations are very informative about the new level of
$d_t$ and thus the magnitude of the innovations should increase to
quickly update $d_t$.

To calculate the score of the log-likelihood it is preferable to use
the autoregressive representation (see, for instance,
\citealp{palma:2007}):
$$(1-B)^{d_t}y_t=y_t + \sum_{j=1}^{\infty}\pi_j(d_t)\,
y_{t-j}=\varepsilon_t\ ,$$
where
\begin{equation}
  \label{eq:defpij}
  \pi_j(d_t)=\prod_{k=1}^{j} \frac{k-1-d_t}{k}=
  \frac{-d_t\, \Gamma(j-d_t)}{\Gamma(1-d_t)\,
  \Gamma(j+1)}=\frac{\Gamma(j-d_t)}{\Gamma(-d_t)\, \Gamma(j+1)}\ .
\end{equation}
In practice, only a finite number $n$ of observations is
available. Therefore, we use the approximation
$$y_t = -\pi_1(d_t)\, y_{t-1} - \pi_2(d_t)\, y_{t-2} - \cdots - \pi_m(d_t)\,
y_{t-m} + \varepsilon_t\ ,$$
with $m < n$. 
Then, the $t$-th contribution, $t=1,\ldots,n$, to the log-likelihood
is:
$$l_t(d_t, \sigma^2) = -\frac12\, \log(\sigma^2) -
\frac{1}{2\,\sigma^2} \left( y_t + \sum_{j=1}^{t-1} \pi_j(d_t)y_{t-j}
\right)^2\,$$
and the corresponding score of the predictive likelihood, see
equation~(\ref{eq:Score_pred}), becomes
\begin{equation}
\label{eq:tvid_score}
\nabla_t = - \frac{1}{\sigma^2}\left( y_t + \sum_{j=1}^{t-1}
\pi_j(d_t)\, y_{t-j} \right) \left( \sum_{j=1}^{t-1} \nu_j(d_t)\,
y_{t-j}\right)\ ,
\end{equation}
where
\begin{equation}
\label{eq:tvid_scale}
\nu_j(d_t) = \frac{\partial\pi_j(d_t)}{\partial d_t} = \pi_j(d_t)
\left(-\Psi(j-d_t) + \Psi(1-d_t) + \frac{1}{d_t}\right)\ ,
\end{equation}
with $\Psi(\cdot)=\Gamma^{'}(\cdot)/\Gamma(\cdot)$ representing the
digamma function.
Finally, we find that $S_t$ in equation~(\ref{eq:Fish_inf}) is
\begin{equation}
\label{eq:defst}
S_t = \sigma^2  \left( \sum_{j=1}^{t-1} \nu_j(d_t)\, y_{t-j}\right)^{-2}\ .
\end{equation}
The calculus details for $\nabla_t$ and $S_t$ are reported in the
Appendix.

\subsection{Parameter estimation}
The static parameter vector $\theta=(\omega,\beta,\alpha,\sigma^2)$ of
the TV-FI model can be estimated by maximum likelihood since the
log-likelihood function can be written in closed form as:
$$\hat{L}_n(\theta) = %\frac1n %???
\sum_{t=1}^{n}l_t(\hat{d_t}(\theta),\sigma^2)$$
where $\hat{d_t}(\theta)$ is obtained recursively using the observed data
$y_1,\ldots,y_n$ as (see
equations~(\ref{eq:tvid_rec}),~(\ref{eq:tvid_score}) and~(\ref{eq:defst}))
$$\hat{d}_{t+1}(\theta) = \omega + \beta\ \hat{d}_{t}(\theta) + \alpha\
s_t(\hat{d_t}(\theta),\sigma^2)\ .$$
Note that we need a starting value $\hat{d}_0$ to initialize the
recursion. Finally, the maximum likelihood estimator is given by
$$\hat{\theta}_n = \mathop{\text{argsup}}_{\theta \in \Theta} \hat{L}_n(\theta)$$
where $\Theta$ is a compact parameter set contained in
$\mathbb{R} \times \mathbb{R} \times \mathbb{R} \times \mathbb{R}^{+}. $
In the next Section, via Monte Carlo experiments, we study the finite
sample behavior of the filtered parameter $\hat{d_t}(\theta)$ and the
maximum likelihood estimator.

\section{Some Monte Carlo results}
\label{sec:mc}

\begin{figure}[ht!]
\centering
\begin{tabular}{c}
	\includegraphics[width=11cm]{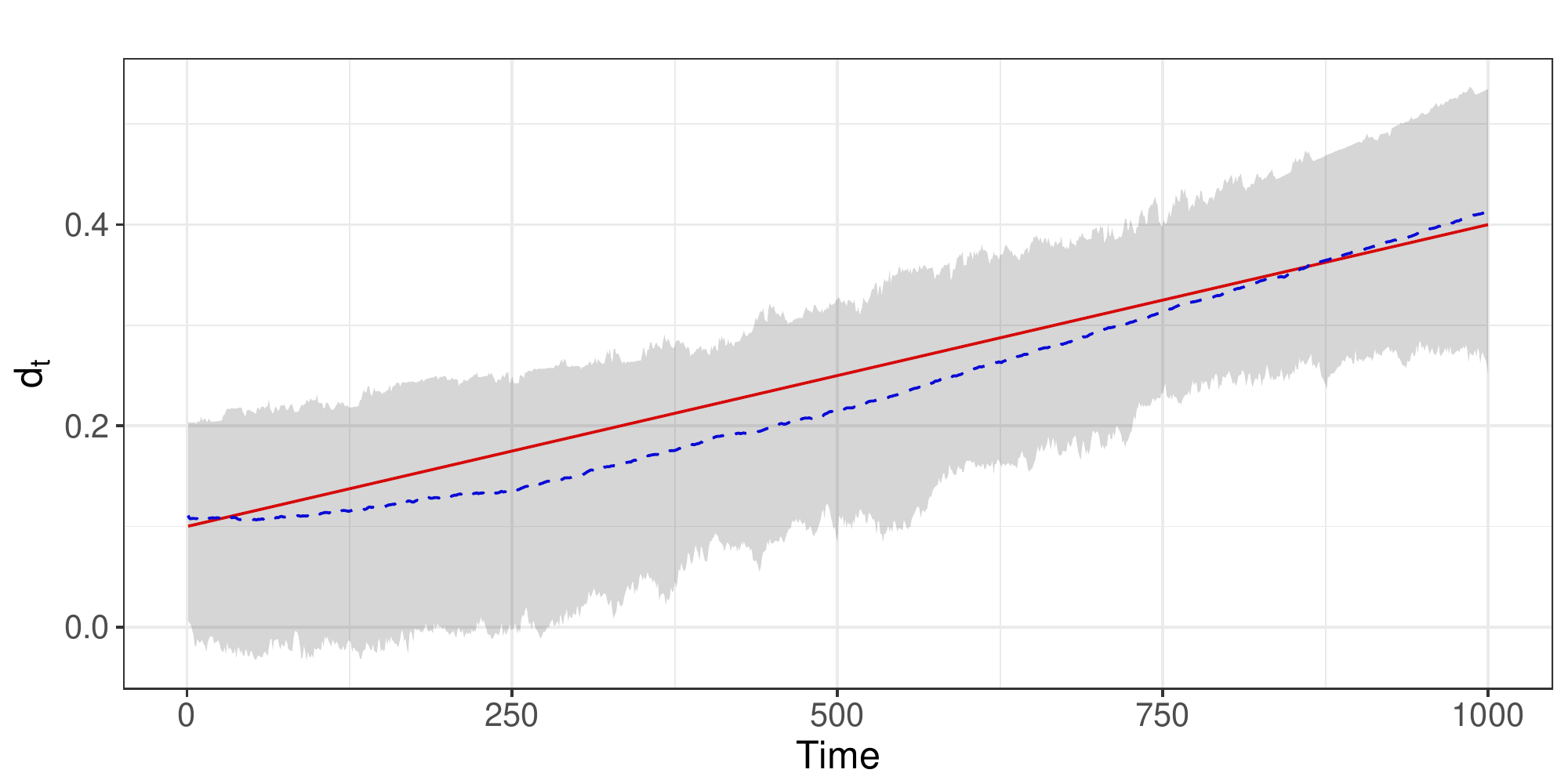} \\
	\includegraphics[width=11cm]{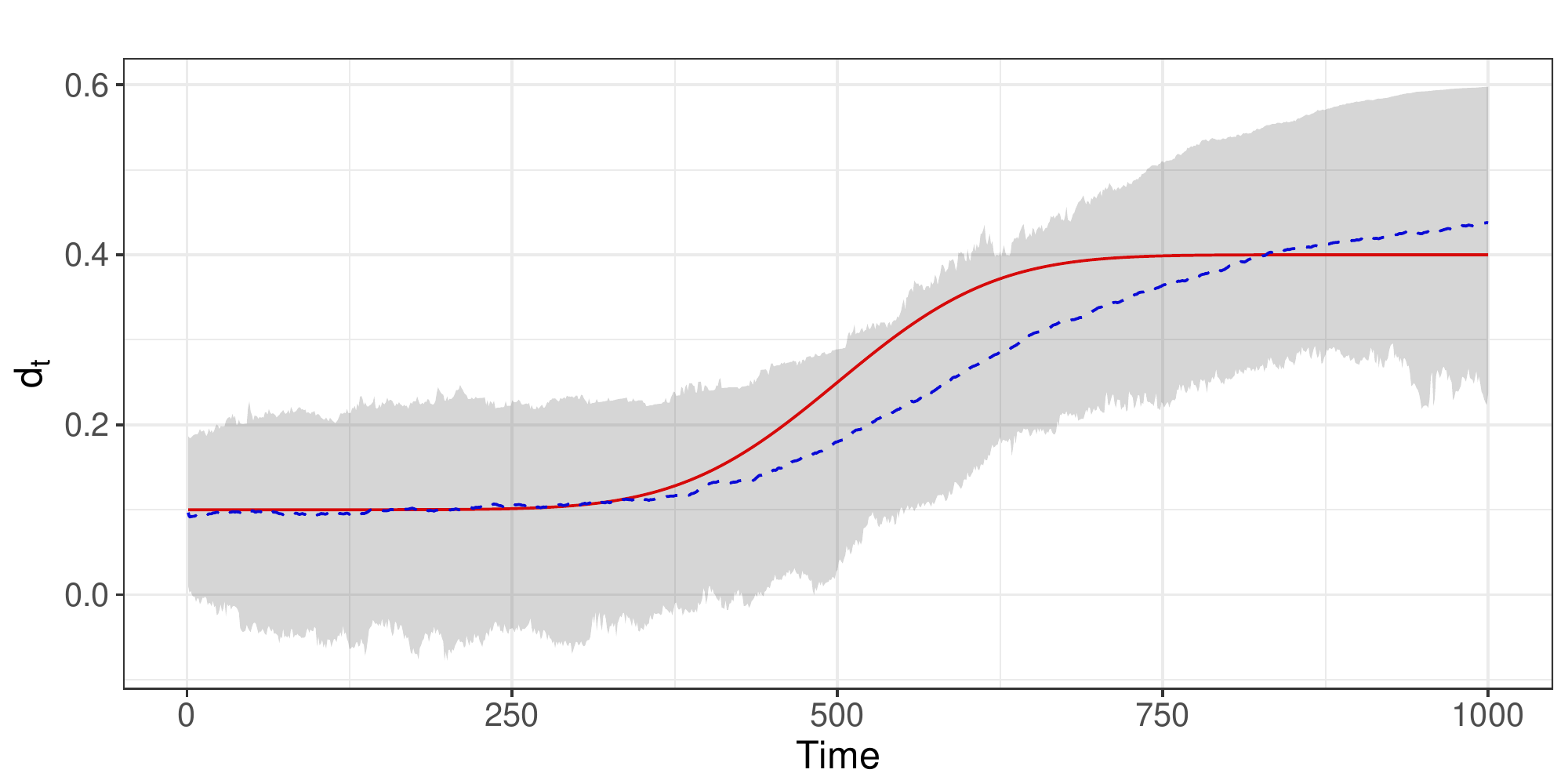}
\end{tabular}

\caption{Result of 200 Monte Carlo simulations, where a time-varying
	fractional parameter (solid line) is estimated with a TV-FI
	model. The dashed line represent the average estimates, while the
	gray band shows the empirical 95\% intervals.} \label{fig:MC}
\end{figure}

In this Section, we carry out some Monte Carlo simulation experiments
in order to establish if the proposed estimation method of the
time-varying long-memory parameter $d$ performs well.

To demonstrate the performance of the proposed method, we simulated
time series data,  $y_1,\ldots,y_n,$ from two TV-FI process:
\begin{equation}
\label{eq:FI}
(1-B)^{d_t}\, y_t = \varepsilon_t\ ,
\end{equation}
where $\varepsilon_t \sim iid \mathcal{N}(0, \sigma^2),$ and $d_t$
is defined, respectively, by:

\begin{equation}
\label{eq:dt_simul_A}
d_t = 0.1 + 0.3\ \frac{t}{n}
\end{equation}
or
\begin{equation}
\label{eq:dt_simul_B}
d_t = 0.1 + 0.3\ \Phi\left(\frac{t-n/2}{3\, \sqrt{n}}\right)\ ,
\end{equation}
with $\Phi(\cdot)$ indicating the standard Gaussian distribution function.

The first specification takes into account a slow increasing trend in
$d_t,$ while the second describes a slow change in regime of $d_t,$
which changes from a short-memory to a persistent situation.

The evolution of $d_t$ is then estimated using the TV-FI model
introduced above. It should be noted that in GAS models the scaling defined 
by~(\ref{eq:Fish_inf}) is often replaced by $S_t^\gamma$,
for some suitable $\gamma$. We found results to be more stable with
$\gamma=0.5$ (see also \citealp{creal:2013}).
Also, GAS models can easily be accommodated in order to
include a link function $\Lambda(\cdot)$, typically with the objective
to constrain the parameter of interest to vary in some region. We used
\[
d_t = \Lambda(g_t) = a + (b-a)\ \frac{e^{g_t}}{1+e^{g_t}}\ ,
\]
so that $d_t \in (a,b)$, while $g_t\in {\rm
	I\!R}$. Recursion~(\ref{eq:tvid_rec}) is then defined in terms of
$g_t$, with~(\ref{eq:tvid_score}) and~(\ref{eq:tvid_scale}) easily
adjusted for the reparametrization.

It should be remarked that $d_0$, the value of the fractional parameter at time
0, is necessary to define the likelihood. We treated $d_0$ as a
parameter to be estimated along with the others.

We obtained 200 Monte Carlo replications from the process defined
by~(\ref{eq:FI}), and~(\ref{eq:dt_simul_A}) or~(\ref{eq:dt_simul_B}),
setting $n=1000$ and $\sigma=2$.

For each replication, the TV-FI model was estimated by maximum
likelihood, setting $(a,b) = (-0.4,0.6)$ and $\omega=0$, while estimating
$(d_0,\alpha,\beta,\sigma)$.

Simulation results are shown in Figure~\ref{fig:MC}. The solid line
shows the true evolution of $d_t$, while the dashed line is its
estimate, averaged over the Monte Carlo replications. The gray band
represents the empirical 95\% intervals.
Figure~\ref{fig:MC} shows how the TV-FI model is able to represent the
evolution of the long-memory parameter, that would be completely
missed by a model with constant $d$.

\section{Empirical application}
\label{sec:empapp}
This Section provides two empirical applications of the proposed
TV-FI model. First we describe the data and then present the
empirical results.

\subsection{Temperature anomalies}
The first data set we analyze is the series of monthly global
historical surface temperature anomalies relative to a 1961--1990
reference period, contained in the data set HadCRUT4. This dataset is
a collaborative product of the Met Office Hadley Centre and the
Climatic Research Unit at the University of East Anglia and contains
data from January, 1850, to August, 2018, on a 5 degree grid, for a
total of $n=2024$ observations (for all details about this dataset see
\citealp{HadCRUT4}).\footnote{The whole dataset can be freely
  downloaded from
  \textit{crudata.uea.ac.uk/cru/data/temperature/\#filfor}}

This series is very interesting since the earth is now in a period of
rising global temperatures and some authors have considered the
stochastic properties of univariate time series of both atmospheric
and oceanic temperatures in an effort to estimate the natural
variability of the earth’s climate. These series often exhibit the
property of statistical long-memory (see \citealp{rea:2011}, and the
references therein). If temperature series are long-memory, the
implications for climatic change are that the temperature series are
mean reverting. In this case, it is possible to support the idea that
the observed rise in global temperatures represents a natural
fluctuation which will reverse in the future.

Actually, the majority of the available studies could not establish the
presence of true long-memory in the temperature series because the
finite sample properties of both long-memory series and series with
structural breaks (\citealp{sibbertsen:2004}). Moreover,
\cite{rea:2011} conclude that none of the temperature series
considered in their paper are true long-memory series, but that the
series are non stationary because of structural changes.

Since changes might concern the fractional parameter $d$, we
think that these are interesting series to apply the model we
propose.

\subsection{Euro-dollar exchange rate}

The second series we consider is the financial time series of the
daily euro-dollar exchange rate from January 1st, 2001, to November 20th,
2018, for a total of $n=4667$ observations \footnote{This dataset
  can be freely downloaded from \textit{finance.yahoo.com}}.

The return series is defined as
$x_t = \ln p_t - \ln p_{t-1}$, $t=2,\ldots n$,
where $p_t$ is the closing quotation of the euro-dollar exchange rate.
Our series of interest is given by the centered absolute returns
$r_t=|x_t-\bar{x}|$, where $\bar{x}$ is the sample mean of $x_t$, as this
series is a good proxy of the volatility. In fact
(\citealp{cotter:2011} and references therein) absolute returns are
robust in the presence of extreme or tail movements; accurate measures
of unobservable latent volatility are obtained from absolute return
volatility asymptotically through the theoretical framework of
realized power variation and, moreover, absolute return volatility
gives desirable finite sample properties that are applicable in
practice for the risk manager.

\subsection{Predictive performance evaluation}

The adequacy of the TV-FI model for the time series at hand is
assessed by evaluating its predictive performance. Since GAS models
are based on parametric assumptions, it is natural to consider
predictions in the form of density forecasts (for reviews on
probabilistic forecasting see e.g.\ \citealp{tai:2000},
\citealp{timmermann:2000}, \citealp{gneiting:2008} and
\citealp{gneiting:2014}). In particular, the
one-step ahead predictive distributions ($h=1$) are analytically
available, while in the multi-step ahead case ($h>1$) they need to be
estimated by simulation. The diagnostic approach used here is the one
based on the maximization of the sharpness of the predictive
distribution, subject to calibration, as proposed
by~\citet{gneiting:2007}. The predictive performance of the TV-FI
model is compared to that of a FI$(d)$ model, with constant $d$, using
proper scoring rules. A popular choice is the continuous ranked probability
score (CRPS), defined as
\[
\int_{-\infty}^{\infty} (F(y) - \mathds{1}\{y \leq z\})^2\, dz\ ,
\]
where $F$ is the predictive CDF \citep{matheson:1976}. Alternative
representations of the CRPS, useful e.g.\ when $F$ is represented by a
sample or when specific regions of interest need to be emphasized,
are discussed in \cite{gneitraft:2007} and \cite{gneitranj:2011}.
Another popular scoring rule is the logarithmic score, which for
the observation $y$ is defined as $-\log(f(y))$, where $f$ is the
predictive density (\citealp{good:1952}; \citealp{bernardo:1979}).
However, this rule lacks robustness (\citealp{selten:1998};
\citealp{gneitraft:2007}), especially for multi-step ahead predictions
($h>1$), when the density (rather than the CDF required by the CRPS)
needs to be estimated, typically with kernel density estimation. The
estimated score may be highly sensitive to the choice of bandwidth,
thus making the ranking of prediction methods more fragile. For these
reasons, the following evaluations will be based on CRPS.

Formal statistical tests of equal predictive performance were also
applied. In particular, we used the \cite{diebold:1995} test
\begin{equation}
\mbox{DM}_l = \sqrt{l}\ \frac{\bar{S}^{TV-FI}_l - \bar{S}^{FI(d)}_l
}{\hat\sigma_l}\ ,
\label{eq:dmtest}
\end{equation}
where $l$ is the length of the out-of-sample period, $\bar{S}^M$
represents the average score for model $M$ and $\hat\sigma^2_l$ is a
suitable estimator of the asymptotic variance of the score difference.
Under the null hypothesis of no difference between the expected scores,
under regularity conditions the test statistic DM$_l$ is
asymptotically standard normal (\citealp{diebold:1995};
\citealp{giacomini:2006}; \citealp{diebold:2015}). Concerning
$\hat\sigma^2_l$, we follow \cite{diks:2011} in using the
heteroskedasticity and autocorrelation consistent (HAC) estimator
defined as
\[
\hat\sigma^2_l = \hat\gamma_0 + 2\,\sum_{j=1}^J \left(1-\frac{j}{J}\right)\,
  \hat\gamma_j\ ,
\]
where $J$ is the largest integer less than or equal to $l^{1/4}$ and
$\hat\gamma_j$ is the lag $j$ sample autocovariance of the sequence
$S_1^{TV-FI}-S_1^{FI(d)},\ldots,S_l^{TV-FI}-S_l^{FI(d)}$ of score
differences over the out-of-sample period.

\subsection{Results for temperature anomalies}

\begin{figure}[ht!]
\centering
\begin{tabular}{c}
  \includegraphics[width=12cm]{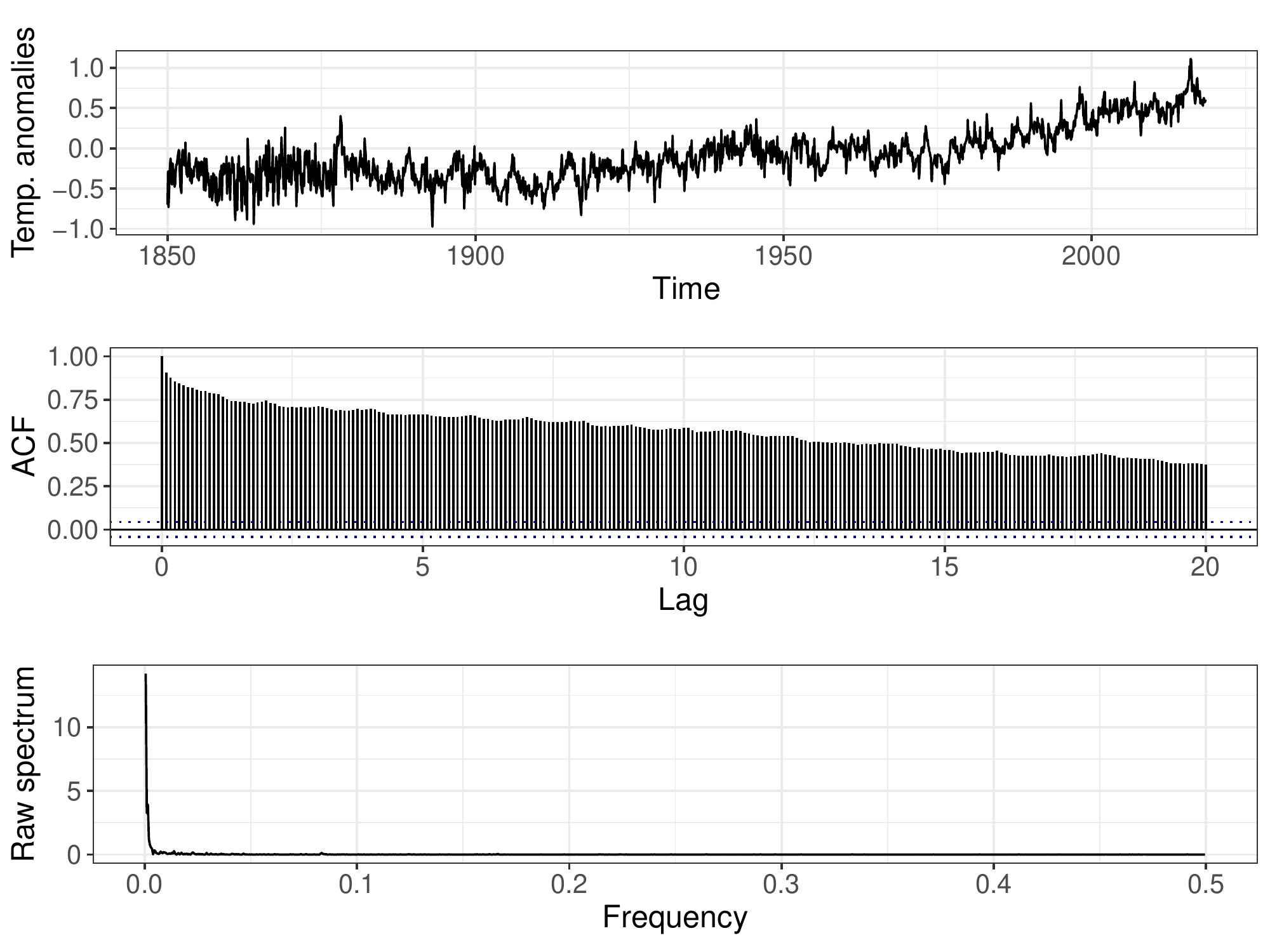}
\end{tabular}

\caption{Temperature anomalies,
  Jan:1950--Aug:2018. Plot of the series, the ACF and the raw
  spectrum.}
\label{fig:anomalies}
\end{figure}

Figure~\ref{fig:anomalies} reports the plot of the series together with
its empirical autocorrelation function and the raw spectrum. From
these plots it is evident that both the slow decaying behavior of the
autocorrelation and the pole near the zero frequency are present, thus
confirming the existence of long-memory behavior. The ADF test rejects
the null hypothesis of unit root; moreover, the maximum likelihood
estimate of the long-memory parameter $d$ calculated for the whole
series is $0.498$, a very high value but lower than $0.5$.
It is also evident that since (about) 1920 the series presents a slow
but increasing trend and that since (about) 1980 the slope of this
trend is greater. We want to investigate whether the model we propose
is able to represent this evolution with a dynamic long-memory
parameter.

\begin{figure}[ht!]
\centering
\begin{tabular}{c}
  \includegraphics[width=12cm]{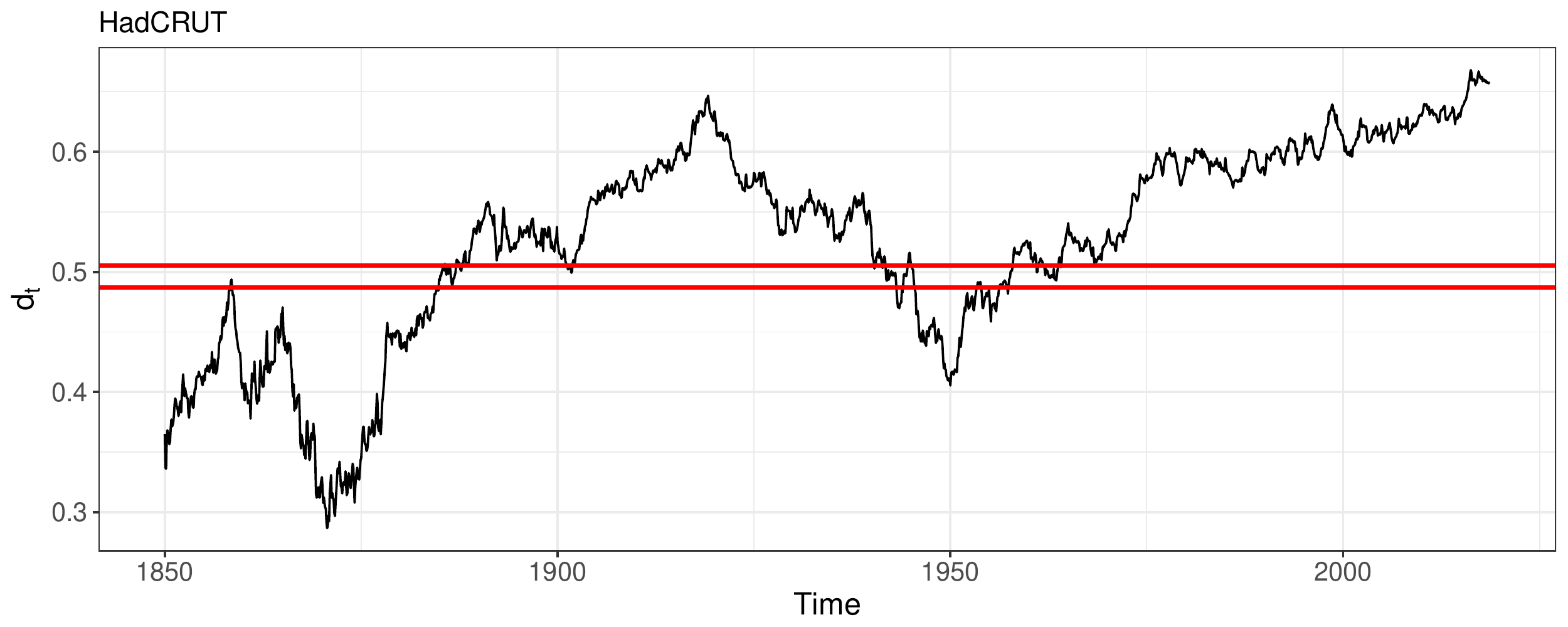}
\end{tabular}

\caption{Temperature anomalies. Evolution of $d_t$
  estimated with the TV-FI model, using the whole series. The
  horizontal band represents the asymptotic confidence interval for a
  constant $d$, also based on the whole series.}
\label{fig:anomalies:dt}
\end{figure}

Figure~\ref{fig:anomalies:dt} reports the results of our estimates,
based on the whole series. In particular, the evolution of $d_t$ is
compared to the asymptotic confidence interval for a constant $d$. It
is possible to see that the evolution of $d_t$ is much greater than
that implied by a constant $d$, with larger values in the second part
of the considered period.

For the predictive performance evaluation, we estimated the TV-FI and
FI$(d)$ models using the first 1000 observations (in-sample
period). For the following out-of-sample period, we computed the
conditional predictive distributions. The TV-FI and FI$(d)$ models are
then compared according to their out-of sample performance, evaluated
on the basis of the CRPS. Every 200 observations, models are
re-estimated (and, therefore, the in-sample period extended).

The average out-of-sample scores for the two models are shown in
Table~\ref{tab:scores:temp}. It should be reminded that models with a
lower score generate more accurate predictions. Hence, the TV-FI model
with dynamic $d$ has a better predictive performance than the FI$(d)$
with constant $d$, and especially so when the prediction horizon $h$
increases.

\begin{table}
\centering
\begin{tabular}{lrrrrrrrrrrrr}\hline
\rule{0em}{1.2em}Prediction horizon $h$ & 1   & 2       & 3       & 6       & 9      & 12     \\\hline
\rule{0em}{1.2em}Average CRPS: TV-FI & 0.0575 &  0.0650 &  0.0709 &  0.0806 & 0.0870 & 0.0910 \\ 
Average CRPS: FI$(d)$                & 0.0583 &  0.0664 &  0.0724 &  0.0830 & 0.0898 & 0.0946 \\ 
DM test                              & -2.253 & -2.640  & -2.365  & -2.709  & -2.858 & -3.442 \\ 
$p$-value                            & 0.012  &  0.004  &  0.009  &  0.003  &  0.002 &  0.000 \\ \hline
  \end{tabular}
  \caption{Temperature anomalies. Comparison of the predictive
    performances for the TV-FI and FI$(d)$ models across the
    out-of-sample period.}
  \label{tab:scores:temp}
\end{table}

As can be seen from Table~\ref{tab:scores:temp}, the DM
test~(\ref{eq:dmtest}) shows that our model with dynamic long-memory
coefficient yields a significant improvement ($p$-values are for a one
directional alternative) in the predictive performance, especially for
longer prediction horizons.

\begin{figure}
  \centering
  \includegraphics[width=12cm]{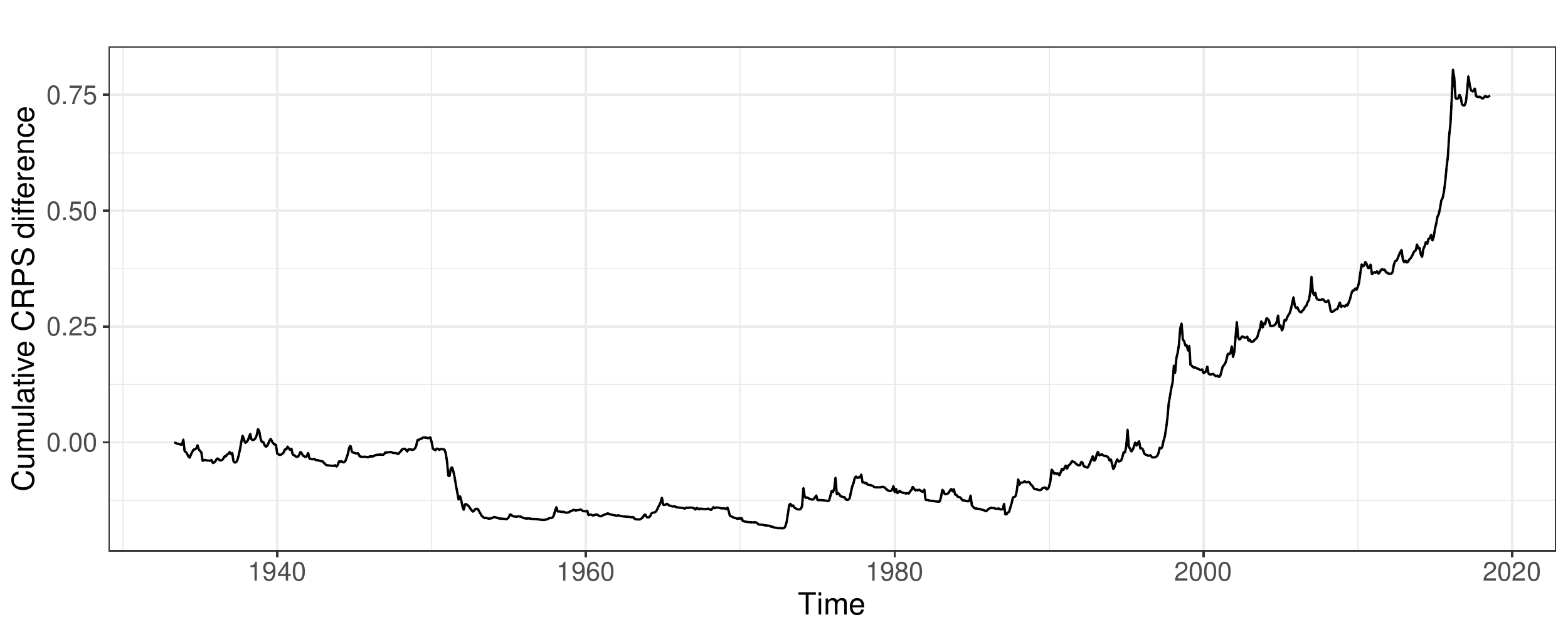}

  \caption{Temperature anomalies. Cumulative difference, in the
    out-of-sample period, between one-step prediction CRPS for the
    FI$(d)$ and TV-FI models.}
  \label{fig:ccrpsd:temp}
\end{figure}

Figure~\ref{fig:ccrpsd:temp} sheds more light on this result, by
showing the evolution, over the out-of-sample period, of the
cumulative sum of the differences between the one-step prediction CRPS 
for the FI$(d)$ and TV-FI models (CS):
\begin{equation}
  \label{eq:ccrpsd}
  \mbox{CS}_j = \sum_{i=1}^j
  \left(S^{FI(d)}_i-S_i^{TV-FI}\right)\ ,\qquad j=1,\ldots,l,
\end{equation}
where $S_i$ is the score for the $i$-th one-step prediction and $l$ is
the length of the out-of-sample period. In
Figure~\ref{fig:ccrpsd:temp}, periods when the TV-FI yields a more
accurate forecast are represented by an upward slope. Interestingly,
it can be seen that the TV-FI model with dynamic long-memory
parameter outperforms the FI$(d)$ model after 1990, i.e.\ when an
increase in the slope of temperature anomalies is observed.

\subsection{Results for the euro-dollar exchange rate}

\begin{figure}[ht!]
\centering
\begin{tabular}{c}
  \includegraphics[width=12cm]{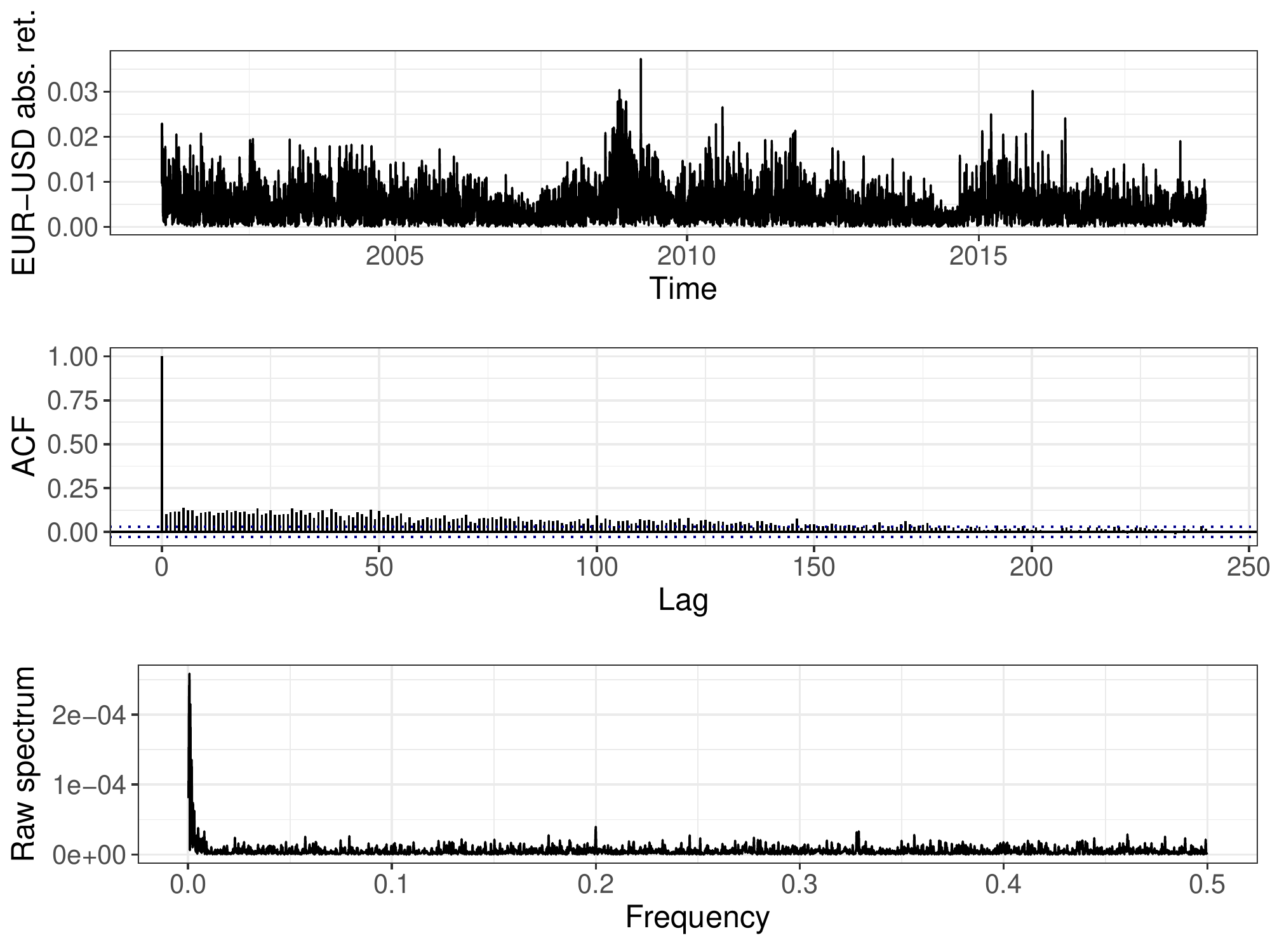} \\
\end{tabular}

\caption{Daily centered absolute log-returns of Euro-dollar exchange rates,
  2001-01-01--2018-11-20. Plots of the series, the ACF and the raw
  spectrum.}
\label{fig:eurusd}
\end{figure}

We report in Figure~\ref{fig:eurusd} the observed series together with
its empirical autocorrelation function and the raw spectrum.
Even if the behavior of this series is completely different from the
previous one, it is possible to see that also this series is
characterized by the qualitative features typical of long-memory
processes. The ADF test rejects the null hypothesis of unit root;
moreover, the maximum likelihood estimate of the long-memory parameter
$d$ calculated for the whole series is $0.131$, indicating that the
effect of shocks is persistent over time.

\begin{figure}[ht!]
\centering
\begin{tabular}{c}
  \includegraphics[width=12cm]{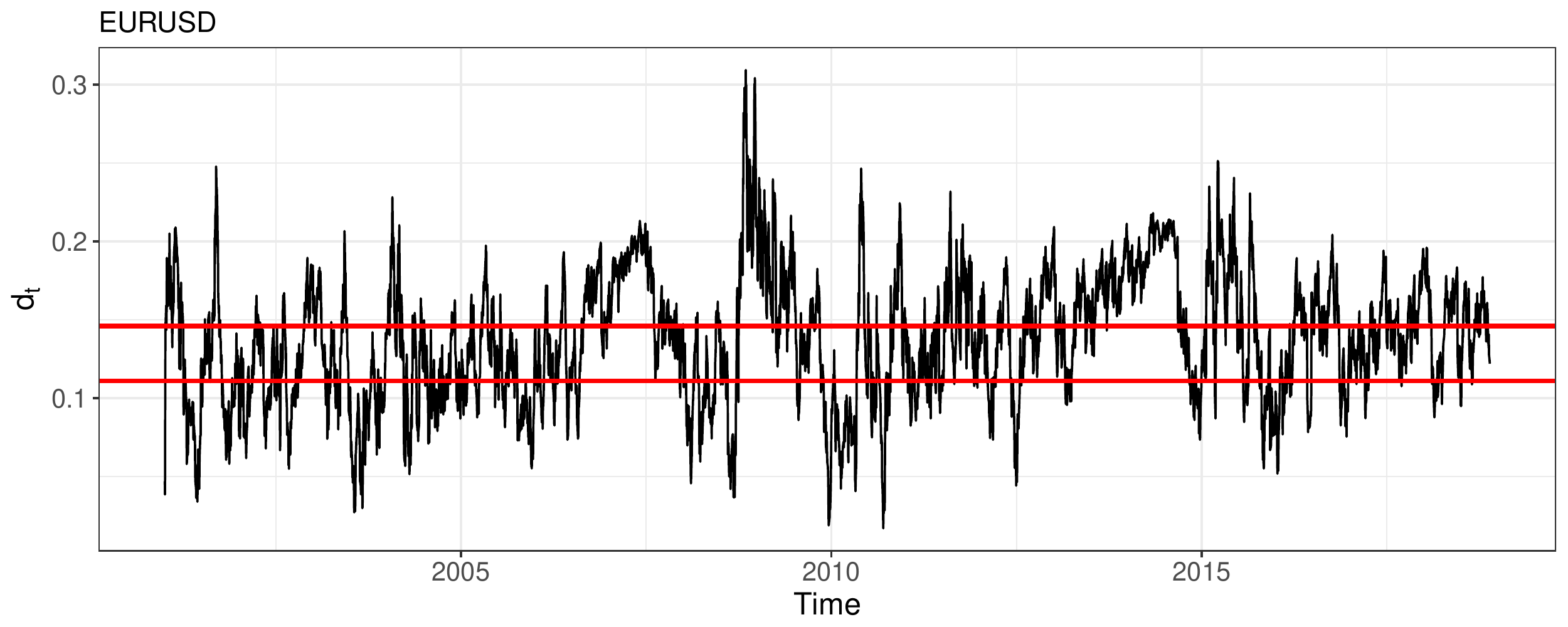}
\end{tabular}

\caption{Centered absolute returns for the euro-dollar exchange
  rate. Evolution of $d_t$ estimated with the TV-FI model, using the
  whole series. The horizontal band represents the asymptotic
  confidence interval for a constant $d$, also based on the whole
  series.}
\label{fig:eurusd:dt}
\end{figure}

Figure~\ref{fig:eurusd:dt} reports the estimated evolution of $d_t$,
compared to to the asymptotic confidence interval for a constant $d$, both
based on the whole series. We see that the TV-FI model implies several
periods in which the estimated $d_t$ remains above the asymptotic
confidence interval.

Concerning the predictive performance, as for temperature anomalies
the conditional predictive distribution are computed after estimating
the TV-FI and FI$(d)$ models with the first 1000 observations
(in-sample period) and updating model estimates every 200 observations.

The average out-of-sample scores for the two models are shown in
Table~\ref{tab:scores:eurusd}. We see that only for $h=1$ the TV-FI
and FI$(d)$ models have the same performance, while for $h>1$  using a
dynamic $d$ improves significantly the predictive performance. This
improvement increases with the prediction horizon $h$.

\begin{table}
\centering
\begin{tabular}{lrrrrrrrrrrrr}\hline
\rule{0em}{1.2em}Prediction horizon $h$ & 1   & 2       & 3       & 6       & 9      & 12     \\\hline
\rule{0em}{1.2em}Average CRPS: TV-FI & 2.0920 &  2.0846 &  2.0874 &  2.0938 & 2.0974 & 2.1003 \\ 
Average CRPS: FI$(d)$                & 2.0998 &  2.1031 &  2.1040 &  2.1154 & 2.1188 & 2.1243 \\ 
DM test                              & -1.127 & -2.593  & -2.577  & -3.416  & -3.892 & -4.616 \\ 
$p$-value                            & 0.130  &  0.005  &  0.005  &  0.000  &  0.000 &  0.000 \\ \hline
  \end{tabular}
  \caption{Centered absolute returns for the euro-dollar exchange
  rate. Comparison of the predictive performances for the TV-FI and
  FI$(d)$ models across the out-of-sample period (the average CRPS is
  multiplied by 1000 to facilitate comparison).}
  \label{tab:scores:eurusd}
\end{table}

\begin{figure}
  \centering
  \includegraphics[width=12cm]{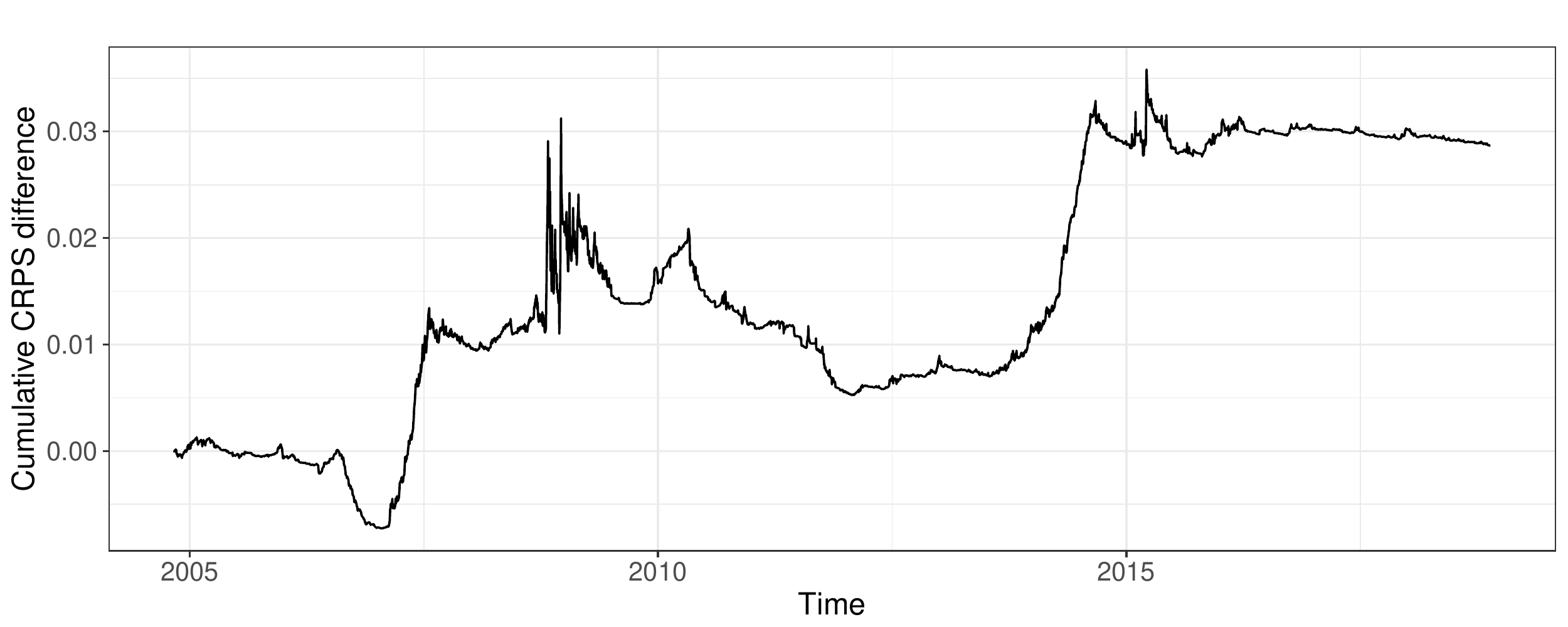}

  \caption{Centered absolute returns for the euro-dollar exchange
  rate. Cumulative difference, in the out-of-sample period, between
  one-step prediction CRPS for the FI$(d)$ and TV-FI models.}
  \label{fig:ccrpsd:eurusd}
\end{figure}

Figure~\ref{fig:ccrpsd:eurusd} shows the evolution of CS$_j$, defined in
equation~(\ref{eq:ccrpsd}) over the out-of-sample period. Upward
slopes represent periods in which the TV-FI model outperforms the the
FI$(d)$ model. Hence, after 2007 and after 2014 we see that the TV-FI
model forecasts more accurately. Interestingly, these are the periods in
which a sharp increase in the volatility of the EUR-USD absolute
returns is observed.

\section{Conclusions}
\label{sec:concl}

In this work we proposed a flexible time-varying fractionally
integrated model which allows the long-memory parameter to vary
dynamically over time. This model is based on the theory of
Generalized Autoregressive Score (GAS) models by \cite{creal:2013} and
\cite{harvey:2013}. The results we obtain are very promising for both
simulated and real time series. In this work we consider only an FI
model but future research may include the extension to a general
ARFIMA$(p,d,q)$ even if, in our opinion, the varying long-memory
parameter, $d_t$, is able to take into account also short memory
components if present.

There are several future directions of research that could improve on
the current work. Missing observations are often present
in empirical studies, for example because unequally spaced time series
are being considered or because stock prices are not recorded during holidays,
despite the underlying values being changing due to external
events. No simple solutions are available for missing observations in
observations-driven models, like the GAS model considered
here. However, the present work could be extended to a context with
missing values by considering the results in~\cite{blasques:2018b}, who use an
indirect inference method to replicate the generating process of the
time series. A different direction of research could concern the
availability of intraday high-frequency data, that has led to the use of
realized variance measures to improve the forecast of volatility,
which is traditionally based, as in the present paper, on
transformations of daily returns. Since realized variance measures are
characterized by even stronger long memory than, e.g., squared daily returns
(\citealp{andersen:2001}), it would be interesting to explore how this
can be exploited in our modeling framework, possibly also using
fractional integrated score dynamics as in \cite{lucas:2016}. From a
more theoretical perspective, a closer link needs to be created with
results on the consistency and asymptotic normality of maximum
likelihood estimators for GAS models (\citealp{blasques:2014b,
  blasques:2014a, blasques:2018a}).

\section{Appendix}
%\subsection{Derivatives of $\nabla_t$ and $S_t$}
Using equation~(\ref{eq:defpij}), we find
\begin{align*}
\nu_j(d_t) &=\frac{\partial \pi_j(d_t)}{\partial d_t}\\
           &= -d_t\
               \frac{-\Gamma'(j-d_t)\, \Gamma(1-d_t)\, \Gamma(j+1) +
                           \Gamma'(1-d_t)\, \Gamma(j+1)\, \Gamma(j-d_t)}
                          {\left(\Gamma(1-d_t)\,\Gamma(j+1)\right)^2}\\
           &-\frac{\Gamma(j-d_t)}{\Gamma(1-d_t)\,\Gamma(j+1)}\\
           &= \frac{-d_t\,\Gamma(j-d_t)}{\Gamma(1-d_t)\,\Gamma(j+1)}
              \left(\frac{-\Gamma'(j-d_t)}{\Gamma(j-d_t)}+\frac{\Gamma'(1-d_t)}{\Gamma(1-d_t)}+\frac{1}{d_t}\right)\\
           &=\pi_j(d_t)\left(-\Psi(j-d_t) + \Psi(1-d_t) +
             \frac{1}{d_t}\right)\ ,
\end{align*}
where $\Psi(\cdot)=\Gamma^{'}(\cdot)/\Gamma(\cdot)$ is the digamma function.
Therefore:
\begin{align*}
\nabla_t &= - \frac{1}{\sigma^2}\left( y_t + \sum_{j=1}^{t-1}
\pi_j(d_t)\, y_{t-j} \right) \left( \sum_{j=1}^{t-1} \frac{\partial \pi_j(d_t)}{\partial d_t}\,
y_{t-j}\right) \\
&=- \frac{1}{\sigma^2}\left( y_t + \sum_{j=1}^{t-1}
\pi_j(d_t)\, y_{t-j} \right) \left( \sum_{j=1}^{t-1} \nu_j(d_t)\,
y_{t-j}\right).
\end{align*}
Now, observe that:
\begin{align*}
\frac{\partial^2 \pi_j(d_t)}{\partial d_t} &=
\nu_j(d_t) \left[ -\Psi(j-d_t) + \Psi(1-d_t) + \frac{1}{d_t}\right] \\
&+ \pi_j(d_t) \left[\Psi^{'}(j-d_t) - \Psi^{'}(1-d_t) -
                                                                          \frac{1}{d^2_t}\right]\ .  
\end{align*}
Hence, we find
\begin{align*}
\mathcal{I}_{t-1} = -E_{t-1}\!\!\left[\frac{\partial \nabla_t}{\partial d_t} \right]
&= \frac{1}{\sigma^2}\, E_{t-1}\!\!\left[ \left( \sum_{j=1}^{t-1}
  \nu_j(d_t)\, y_{t-j} \right)^2 \right.\\ 
&+ \left.\left( y_t + \sum_{j=1}^{t-1}\pi_j(d_t)\, y_{t-j}\right)
\left( \sum_{j=1}^{t-1} \frac{\partial^2 \pi_j(d_t)}{\partial d_t}\,
                                              y_{t-j}\right)   \right]\\
&= \frac{1}{\sigma^2}\, \left( \sum_{j=1}^{t-1}
  \nu_j(d_t)\, y_{t-j} \right)^2\ ,
\end{align*}
where we used $E_{t-1}\!\left[y_t +
  \sum_{j=1}^{t-1}\pi_j(d_t)\, y_{t-j}\right]=E_{t-1}\!\left[\epsilon_t\right]=0$.
Finally:
$$S_t=\mathcal{I}_{t-1}^{-1} = \sigma^2\, \left( \sum_{j=1}^{t-1}
  \nu_j(d_t)\, y_{t-j} \right)^{-2}\ .$$
%======================================================================
\bibliographystyle{chicago}
\bibliography{biblio_GAS}

\begin{thebibliography}{}

\bibitem[\protect\citeauthoryear{Andersen, Bollerslev, Diebold, and
  Labys}{Andersen et~al.}{2001}]{andersen:2001}
Andersen, T.~G., T.~Bollerslev, F.~X. Diebold, and P.~Labys (2001).
\newblock The distribution of realized exchange rate volatility.
\newblock {\em Journal of the American Statistical Association\/}~{\em
  96\/}(453), 42--55.

\bibitem[\protect\citeauthoryear{Beran and Terrin}{Beran and
  Terrin}{1996}]{beran:1996}
Beran, J. and N.~Terrin (1996).
\newblock Testing for a change of the long-memory parameter.
\newblock {\em Biometrika\/}~{\em 83}, 627--638.

\bibitem[\protect\citeauthoryear{Bernardo}{Bernardo}{1979}]{bernardo:1979}
Bernardo, J. (1979).
\newblock Expected information as expected utility.
\newblock {\em The Annals of Statistics\/}~{\em 7\/}(3), 686--690.

\bibitem[\protect\citeauthoryear{Blasques, Gorgi, and Koopman}{Blasques
  et~al.}{2018}]{blasques:2018b}
Blasques, F., P.~Gorgi, and S.~J. Koopman (2018).
\newblock Missing observations in observation-driven time series models.
\newblock Tinbergen Institute Discussion Papers, 2018-013/III.

\bibitem[\protect\citeauthoryear{Blasques, Gorgi, Koopman, and
  Wintenberger}{Blasques et~al.}{2018}]{blasques:2018a}
Blasques, F., P.~Gorgi, S.~J. Koopman, and O.~Wintenberger (2018).
\newblock Feasible invertibility conditions and maximum likelihood estimation
  for observation-driven models.
\newblock {\em Electronic Journal of Statistics\/}~{\em 12\/}(1), 1019--1052.

\bibitem[\protect\citeauthoryear{Blasques, Koopman, and Lucas}{Blasques
  et~al.}{2014a}]{blasques:2014b}
Blasques, F., S.~J. Koopman, and A.~Lucas (2014a).
\newblock Maximum likelihood estimation for correctly specified generalized
  autoregressive score models: Feedback effects, contraction conditions and
  asymptotic properties.
\newblock Tinbergen Institute Discussion Papers, 14-074/III.

\bibitem[\protect\citeauthoryear{Blasques, Koopman, and Lucas}{Blasques
  et~al.}{2014b}]{blasques:2014a}
Blasques, F., S.~J. Koopman, and A.~Lucas (2014b).
\newblock Maximum likelihood estimation for score-driven models.
\newblock Tinbergen Institute Discussion Papers, 14-029/III.

\bibitem[\protect\citeauthoryear{Boubaker}{Boubaker}{2018}]{boubaker:2018}
Boubaker, H. (2018).
\newblock A generalized arfima model with smooth transition fractional
  integration parameter.
\newblock {\em Journal of Time Series Econometrics\/}~{\em 10}, 1--20.

\bibitem[\protect\citeauthoryear{Boutahar, Dufr\'{e}not, and
  P\'{e}guin-Feissolle}{Boutahar et~al.}{2008}]{boutahar:2008}
Boutahar, M., G.~Dufr\'{e}not, and A.~P\'{e}guin-Feissolle (2008).
\newblock A simple fractionally integrated model with a time-varying long
  memory parameter $d_t$.
\newblock {\em Computational Economics\/}~{\em 31}, 225--241.

\bibitem[\protect\citeauthoryear{Caporin and Pres}{Caporin and
  Pres}{2013}]{Caporin:2013}
Caporin, M. and J.~Pres (2013).
\newblock Forecasting temperature indeces density with time-varying long-memory
  models.
\newblock {\em Journal of Forecasting\/}~{\em 32}, 339--352.

\bibitem[\protect\citeauthoryear{Cotter}{Cotter}{2011}]{cotter:2011}
Cotter, J. (2011).
\newblock Absolute return volatility.
\newblock Working Papers 200415, Geary Institute, University College Dublin.

\bibitem[\protect\citeauthoryear{Creal, Koopman, and Lucas}{Creal
  et~al.}{2013}]{creal:2013}
Creal, D., S.~Koopman, and A.~Lucas (2013).
\newblock Generalized autoregressive score models with applications.
\newblock {\em Journal of Applied Econometrics\/}~{\em 28}, 777--795.

\bibitem[\protect\citeauthoryear{Diebold}{Diebold}{2015}]{diebold:2015}
Diebold, F.~X. (2015).
\newblock Comparing predictive accuracy twenty years later: A personal
  perspective on the use and abuse of diebold-mariano tests.
\newblock {\em Journal of Business \& Economic Statistics\/}~{\em 33\/}(1),
  1--9.

\bibitem[\protect\citeauthoryear{Diebold and Mariano}{Diebold and
  Mariano}{1995}]{diebold:1995}
Diebold, F.~X. and R.~S. Mariano (1995).
\newblock Comparing predictive accuracy.
\newblock {\em Journal of Business \& Economic Statistics\/}~{\em 13\/}(3),
  253--263.

\bibitem[\protect\citeauthoryear{Diks and Van~Dijk}{Diks and
  Van~Dijk}{2011}]{diks:2011}
Diks, Cees, P.~V. and D.~Van~Dijk (2011).
\newblock Likelihood-based scoring rules for comparing density forecasts in
  tails.
\newblock {\em Journal of Econometrics\/}~{\em 163\/}(2), 215--230.

\bibitem[\protect\citeauthoryear{Giacomini and White}{Giacomini and
  White}{2006}]{giacomini:2006}
Giacomini, R. and H.~White (2006).
\newblock Tests of conditional predictive ability.
\newblock {\em Econometrica\/}~{\em 74\/}(6), 1545--1578.

\bibitem[\protect\citeauthoryear{Gneiting}{Gneiting}{2008}]{gneiting:2008}
Gneiting, T. (2008).
\newblock Editorial: Probabilistic forecasting.
\newblock {\em Journal of the Royal Statistical Society Series A\/}~{\em
  171\/}(2), 319--321.

\bibitem[\protect\citeauthoryear{Gneiting, Balabdaoui, and Raftery}{Gneiting
  et~al.}{2007}]{gneiting:2007}
Gneiting, T., F.~Balabdaoui, and A.~E. Raftery (2007).
\newblock Probabilistic forecasts, calibration and sharpness.
\newblock {\em Journal of the Royal Statistical Society Series B\/}~{\em
  69\/}(2), 243--268.

\bibitem[\protect\citeauthoryear{Gneiting and Katzfuss}{Gneiting and
  Katzfuss}{2014}]{gneiting:2014}
Gneiting, T. and M.~Katzfuss (2014).
\newblock Probabilistic forecasting.
\newblock {\em Annual review of statistics and its application\/}~{\em 1},
  125--151.

\bibitem[\protect\citeauthoryear{Gneiting and Raftery}{Gneiting and
  Raftery}{2007}]{gneitraft:2007}
Gneiting, T. and A.~E. Raftery (2007).
\newblock Strictly proper scoring rules, prediction, and estimation.
\newblock {\em Journal of the American Statistical Association\/}~{\em
  102\/}(477), 359--378.

\bibitem[\protect\citeauthoryear{Gneiting and Ranjan}{Gneiting and
  Ranjan}{2011}]{gneitranj:2011}
Gneiting, T. and R.~Ranjan (2011).
\newblock Comparing density forecasts using threshold- and quantile-weighted
  scoring rules.
\newblock {\em Journal of Business \& Economic Statistics\/}~{\em 29\/}(3),
  411--422.

\bibitem[\protect\citeauthoryear{Good}{Good}{1852}]{good:1952}
Good, I. (1852).
\newblock Rational decisions.
\newblock {\em Journal of the Royal Statistical Society Series B\/}~{\em
  14\/}(1), 107--114.

\bibitem[\protect\citeauthoryear{Granger and Joyeux}{Granger and
  Joyeux}{1980}]{granger:1980}
Granger, C. and R.~Joyeux (1980).
\newblock An introduction to long-range time series models and fractional
  differencing.
\newblock {\em Journal of Time Series Analysis\/}~{\em 1}, 15--30.

\bibitem[\protect\citeauthoryear{Harvey}{Harvey}{2013}]{harvey:2013}
Harvey, A. (2013).
\newblock {\em Dynamic Models for Volatility and Heavy Tails: With Applications
  to Financial and Economic Time Series}.
\newblock Cambridge: University Press.

\bibitem[\protect\citeauthoryear{Hassler and Meller}{Hassler and
  Meller}{2014}]{hasller:2014}
Hassler, U. and B.~Meller (2014).
\newblock Detecting multiple breaks in long memory the case of u.s. inflation.
\newblock {\em Empirical Economics\/}~{\em 46}, 653--680.

\bibitem[\protect\citeauthoryear{Hosking}{Hosking}{1981}]{hosking:1981}
Hosking, J. (1981).
\newblock Fractional differencing.
\newblock {\em Biometrika\/}~{\em 68}, 165--176.

\bibitem[\protect\citeauthoryear{Jensen and Whitcher}{Jensen and
  Whitcher}{2000}]{jensen2000}
Jensen, M.~J. and B.~Whitcher (2000).
\newblock Time-varying long-memory in volatility: Detection and estimation with
  wavelets.

\bibitem[\protect\citeauthoryear{Lu and Guegan}{Lu and Guegan}{2011}]{lu:2011}
Lu, Z. and D.~Guegan (2011).
\newblock Estimation of time-varying long memory parameter using wavelet
  method.
\newblock {\em Communications in Statistics - Simulation and
  Computation\/}~{\em 40}, 596--613.

\bibitem[\protect\citeauthoryear{Lucas and Opschoor}{Lucas and
  Opschoor}{2016}]{lucas:2016}
Lucas, A. and A.~Opschoor (2016).
\newblock Fractional integration and fat tails for realized covariance kernels
  and returns.
\newblock Tinbergen Institute Discussion Papers, 2016-069/IV.

\bibitem[\protect\citeauthoryear{Matheson and Winkler}{Matheson and
  Winkler}{1976}]{matheson:1976}
Matheson, J.~E. and R.~L. Winkler (1976).
\newblock Scoring rules for continuous probability distributions.
\newblock {\em Management Science\/}~{\em 22\/}(10), 1087--1096.

\bibitem[\protect\citeauthoryear{Morice, Kennedy, Rayner, and P.D.}{Morice
  et~al.}{2012}]{HadCRUT4}
Morice, C., J.~Kennedy, N.~Rayner, and J.~P.D. (2012).
\newblock Quantifying uncertainties in global and regional temperature change
  using an ensemble of observational estimates: The hadcrut4 dataset.
\newblock {\em Journal of Geophysical Research\/}~{\em 117}, 1--22.

\bibitem[\protect\citeauthoryear{Palma}{Palma}{2007}]{palma:2007}
Palma, W. (2007).
\newblock {\em Long-memory time series}.
\newblock New Jersey: Wiley.

\bibitem[\protect\citeauthoryear{Ray and Tsay}{Ray and Tsay}{2002}]{ray:2002}
Ray, B. and R.~Tsay (2002).
\newblock Bayesian methods for change-point detection in long-range dependent
  processes.
\newblock {\em Journal of Time Series Analysis\/}~{\em 23}, 687--705.

\bibitem[\protect\citeauthoryear{Rea, Reale, and J.}{Rea
  et~al.}{2011}]{rea:2011}
Rea, W., M.~Reale, and B.~J. (2011).
\newblock Long memory in temperature reconstructions.
\newblock {\em Climatic Change\/}~{\em 107}, 247--265.

\bibitem[\protect\citeauthoryear{Roueff and von Sachs}{Roueff and von
  Sachs}{2011}]{roueff:2011}
Roueff, F. and R.~von Sachs (2011).
\newblock Locally stationary long memory estimation.
\newblock {\em Stochastic Processes and their Applications\/}~{\em 121},
  813--844.

\bibitem[\protect\citeauthoryear{Selten}{Selten}{1998}]{selten:1998}
Selten, R. (1998).
\newblock Axiomatic characterization of the quadratic scoring rule.
\newblock {\em Experimental Economics\/}~{\em 1}, 43--62.

\bibitem[\protect\citeauthoryear{Sibbertsen}{Sibbertsen}{2004}]{sibbertsen:2004}
Sibbertsen, P. (2004).
\newblock Long memory versus structural breaks: an overview.
\newblock {\em Statistical Papers\/}~{\em 45}, 465--515.

\bibitem[\protect\citeauthoryear{Tay and Wallis}{Tay and
  Wallis}{2000}]{tai:2000}
Tay, A.~S. and K.~F. Wallis (2000).
\newblock Density forecasting: A survey.
\newblock {\em Journal of Forecasting\/}~{\em 19\/}(4), 235--254.

\bibitem[\protect\citeauthoryear{Timmermann}{Timmermann}{2000}]{timmermann:2000}
Timmermann, A. (2000).
\newblock Density forecasting in economics and finance.
\newblock {\em Journal of Forecasting\/}~{\em 19\/}(4), 231--234.

\bibitem[\protect\citeauthoryear{Yamaguchi}{Yamaguchi}{2011}]{yamaguchi:2011}
Yamaguchi, K. (2011).
\newblock Estimating a change point in the long memory parameter.
\newblock {\em Journal of Time Series Analysis\/}~{\em 32}, 304--314.

\end{thebibliography}
%======================================================================

\end{document}